\documentclass[twocolumn,aps,prl,showpacs,preprintnumbers,amsmath,amssymb, superscriptaddress]{revtex4}

\usepackage{graphicx}
\usepackage{dcolumn}
\usepackage{bm}

\begin{document}


\title{Percolation through Voids around Randomly Oriented Platonic Solids} 


\author{D. J. Priour, Jr and N. J. McGuigan}
\affiliation{Department of Physics \& Astronomy, Youngstown State University, Youngstown, OH 44555, USA}


\date{\today}

\begin{abstract}
Porous materials made up of impermeable polyhedral grains constrain 
fluid flow to voids around the impenetrable constituent barrier particles.  
A percolation transition marks the boundary between assemblies of grains which 
contain system spanning void networks, admitting bulk transport, and configurations 
which may not be traversed on macroscopic scales.  
With dynamical infiltration of void spaces using virtual tracer particles, 
we give an exact treatment of grain geometries, and
we calculate critical densities  for polyhedral inclusions for the
five platonic solids (i.e. tetrahedra, cubes, octahededra, dodecahedra, and   
icosahedra).  In each case, we calculate percolation threshold concentrations $\rho_{c}$ for 
aligned and randomly oriented grains, finding 
distinct $\rho_{c}$ values for the former versus the latter only for cube-shaped grains.
We calculate the dynamical scaling exponent at the percolation threshold, finding 
subdiffusive value $z = 0.19(1)$ common to all grain shapes considered. 
\end{abstract}
\pacs{64.60.ah,61.43.Gt,64.60.F-}

\maketitle
In a variety of relevant situations, porous materials are comprised of impermeable grains with fluid flow or 
ion transport through networks of voids surrounding the inclusions.  Whereas contiguous 
void spaces form system spanning networks at low densities, void networks on a macroscopic 
scale cease to exist beyond a threshold density of grains.  The latter, marking the 
transition from assemblies of grains allowing bulk fluid flow and those which do not 
is a percolation transition with all the hallmarks of 
a continuous phase transition.   

Grains making up porous media occupy a spectrum of shapes from weathered round geometries to 
angular faceted polyhedral crystallites.  We examine percolation phenomena along this continuum of 
grain shapes by calculating percolation thresholds for configurations of randomly placed polyhedral barriers 
for each of the Platonic solids (i.e. tetrahedra, cubes, octahedra, dodecahedra, and icosahedra).  
To incorporate the strong 
disorder often present in naturally occurring granular media, we calculate percolation thresholds for  
interpenetrating polyhedral inclusions with random orientations.  To our knowledge
our study is the first to calculate percolation thresholds around randomly oriented polyhedra of any type.

Fluid or ion flow through voids around impenetrable barrier particles is a scenario complementary to charge   
transport through contiguous conducting grains, as in metallic particles sintered  into a random
conducting network.  The latter situation is amenable to techniques such as the Hoshen-Kopelman algorithm~\cite{Hoshen} 
for finding clusters of connected grains to determine if a system spanning conducting network exists. However, for 
voids around overlapping grains, cluster infiltration methods are more difficult to implement with interstitial spaces
often not readily partitioned into discrete elements.

Assemblies of highly symmetrical inclusions such as randomly placed spheres 
are amenable to an analysis involving Voronoi tesselation~\cite{Elam,Marck,Rintoul}.  On the other hand,
discretization schemes have been used to extrapolate to the continuum limit in the case of  
randomly placed
spheres~\cite{Martys,Maier}, ellipsoids~\cite{Yi1,Yi2}, and aligned cubes~\cite{Koza}.  To find percolation thresholds for randomly  
oriented polyhedral grains, we circumvent an explicit description of void shapes with an exact treatment of
the geometry of  barrier particles by using virtual tracer particles to 
dynamically infiltrate void spaces. Previous examples of dynamical simulations 
in 3D have been described in the context of randomly placed spheres~\cite{Kammerer,Hofling,Spanner,dynamic}.
For a computationally efficient and physically plausible implementation,
tracer particles propagate from an initial location at the center of 
a large sample of randomly placed inclusions, following
linear trajectories punctuated by specular reflections from the planar surfaces of 
included polyhedra. 

We examine the impact of orientational disorder by calculating critical densities $\rho_{c}$ for
assemblies of grains in which the constituent polyhedra are randomly oriented as well as for 
aligned barriers, finding distinct percolation thresholds for the latter and former scenarios only in the case 
of cubic inclusions; for all other polyhedra under consideration the $\rho_{c}$ values are identical up to  
Monte Carlo statistical error.

As in a recent study~\cite{dynamic2} we calculate
percolation thresholds using two complementary methods.
On the one hand, data collapses in the context of finite size scaling analysis yield
$\rho_{c}$ and the dynamical critical exponent $z$. 
Alternatively, comparing effective dynamical scaling exponents calculated from the RMS displacements 
for distinct time sub-intervals  is a 
means to determine if one is above or below $\rho_{c}$, and thereby find the 
threshold density; agreement among results of the two techniques is comparable to  
the Monte Carlo error.

In order to extrapolate to the thermodynamic limit and find $\rho_{c}$, we average over disorder to suppress 
statistical fluctuations, and consider at least $4 \times 10^{4}$ cubic volumes of randomly placed inclusions. 
For assemblies of grains to be effectively infinite in size, we insist that none of the virtual tracers 
travel far enough to breach the simulation volume.  To this end, we use 
system sizes ranging from $L = 90$ to $L = 170$ in units of the radius of spheres circumscribed about the 
polyhedra; smaller sizes are 
appropriate for the more densely packed angular tetrahedral grains, while larger systems are needed for the bulkier 
dodecahedra and icosahedra.
To allow for at least on the order of $10^{6}$ collisions with faceted inclusions,
tracer dwell times are $2 \times 10^{5}$ to $5 \times 10^{5}$ time units (with normalized inter-collision
velocities where $|\vec{v} | = 1$), depending on the type of barrier particle.

To mitigate the computational burden of locating the nearest polyhedral surface, the cubic simulation volumes 
are partitioned into voxels a unit on a side, with disorder realizations generated voxel by voxel for optimal memory
usage.  The probability for the geometric centers of $n$ grains in a voxel, $P(n) = (\rho v)^{n} e^{-\rho v}/n!$ ($v = 1$ being 
the voxel volume), may be efficiently and accurately sampled with a robust random number generator.  

In identifying collisions of tracer particles with grains, 
neighboring voxels are checked for penetration of spheres circumscribed about polyhedra; in the event of contact with a sphere, each planar facet 
of the polyhedron within is examined 
for an intersection with the tracer trajectory.  

The fact that regular convex polyhedra are 
the set of all points confined within the planes containing the $m$ facets is  
tantamount to the constraint  $\hat{n}_{i} \cdot (\vec{V} - \vec{V}_{0})  \leq \epsilon$ 
with $\vec{V}_{0}$ being the geometric center of the 
solid and $\vec{V}$ a point within the polyhedron; 
$\hat{n}_{i}$ is the unit vector normal to the $i$th facet, and $\epsilon$ the distance of each facet plane to the 
polyhedron center.  For a collision with the $i$th facet, the relation becomes an equality which may be solved for a transit time, 
with the polyhedron constraint
imposed for each of the $m - 1$ remaining faces to reject spurious scattering events.  
Appealing to the geometries of the corresponding polyhedra gives $\epsilon = 1/3$ for the tetrahedron, $\epsilon = 1/\sqrt{3}$ for   
the cube and octahedron, and $\epsilon = [(5 + 2 \sqrt{5})/15]^{1/2}$ for dodecahedra and icosahedra.

\begin{figure}
\includegraphics[width=.45\textwidth]{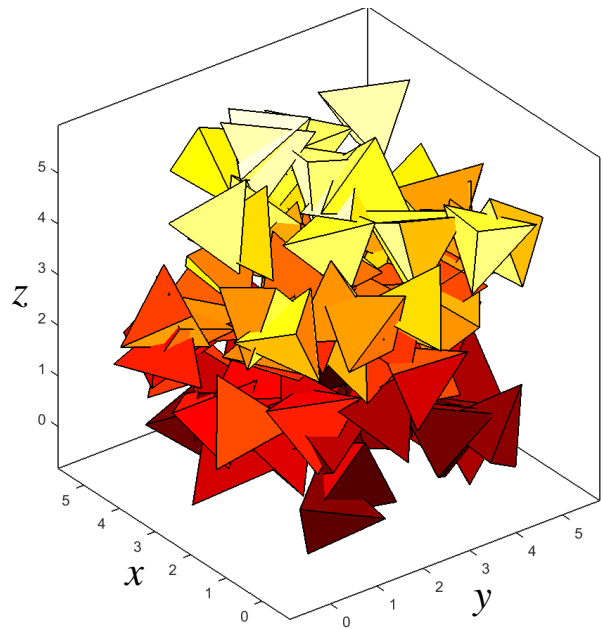}
\caption{\label{fig:Fig1} (Color online) Configurations of randomly oriented tetrahedral grains with a subset (i.e. 100) of the tetrahedra 
shown.}
\end{figure}

\begin{figure}
\includegraphics[width=.45\textwidth]{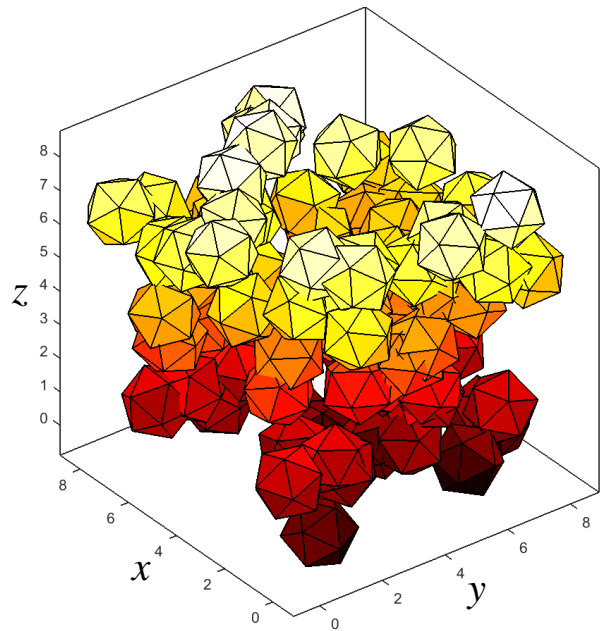}
\caption{\label{fig:Fig2} (Color online) Configurations of randomly oriented icosahedral grains with a subset (i.e. 100) of the 
icosahedra shown.} 
\end{figure}

The implementation of an aligned or randomly oriented scheme for the polyhedral inclusions is determined by the choice of the local coordinate 
system for each grain.  For   
aligned polyhedra, normalized vectors corresponding to each planar facet are specified in relation to a common set of Cartesian 
axes.  On the other hand, for randomly oriented grains, distinct orthogonal coordinate systems are selected   
randomly for each inclusion. To prevent an orientational bias, we 
isotropically choose two normalized vectors $\hat{v}_{1}$ and $\hat{v}_{2}$.  The latter 
are converted into an orthogonal pair by putting $\hat{u}_{1} = \hat{v}_{1}$, with $\hat{u}_{2}$ obtained by normalizing 
$\hat{v}_{2} - (\hat{v}_{1} \cdot \hat{v}_{2} ) \hat{v}_{1}$ to remove the component parallel to $\hat{v}_{1}$.  A third perpendicular axis is 
then fixed with $\hat{u}_{3} = \hat{u}_{1} \times \hat{u}_{2}$.  The components for the vectors $\hat{n}_{i}^{'}$ normal to the facets of a 
polyhedron are hence $n_{ik}^{'} = \sum_{j=1}^{3} n_{ij} u_{jk}$ with $j$  and $k$ corresponding to the randomly oriented and common 
coordinate systems, respectively.  Assemblies of randomly oriented polyhedra are shown in Fig.~\ref{fig:Fig1} for tetrahedra and
in Fig.~\ref{fig:Fig2} for icosahedra. 

\begin{figure}
\includegraphics[width=.4\textwidth]{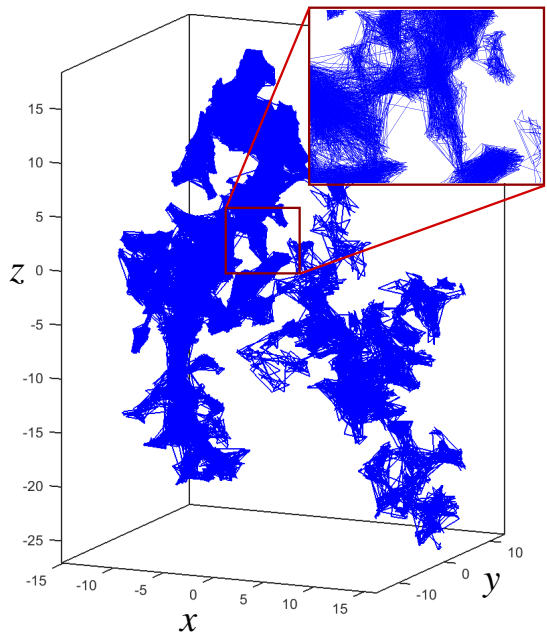}
\caption{\label{fig:Fig3} (Color online) Trajectories for tracer particles moving among 
randomly oriented icosahedral grains are show for $\rho$ just below $\rho_{c}$ for $4 \times 10^{6}$ time units.
Blue segments represent 400 collision events, and the boxed image in the upper right hand corner is a magnified view 
of the indicated section of the void network.} 
\end{figure}

A trajectory sequence for a tracer particle passing through void networks among randomly oriented icosahedral grains is 
shown in Fig.~\ref{fig:Fig3}, with each blue segment representing 400 collisions for the sake of clarity  
in highlighting void networks.  In the red box is an expanded view showing the narrow regions joining
larger void spaces.

Critical indices, such as the threshold density $\rho_{c}$ and the dynamical critical exponent $z$ may be calculated using the framework of 
single parameter finite size scaling theory~\cite{Stauffer}; with systems effectively infinite in size,
RMS displacements are calculated for a range of finite times. 
In terms of elapsed times (we consider 16 time subintervals in this work) rather than sizes,
one anticipates a scaling form $\delta_{\mathrm{RMS}} (\rho) = t^{z} f[t^{y} (\rho - \rho_{c} )]$~\cite{Stauffer},
where $z$ is the dynamical critical exponent and $y$ a scaling exponent. 

Plotting $t^{-z} f$ with respect to $t^{y} ( \rho - \rho_{c})$ for $\rho$ near $\rho_{c}$ in principle yields a data collapse
with Monte Carlo data falling on a single curve.  We use the data collapse phenomenon as a 
quantitative tool by considering a scaling function $g(x) = \sum_{j = 0}^{n} A_{j} x^{j}$, and optimizing with respect to the $A_{j}$ coefficients as 
well as $\rho_{c}$, $z$, and $y$ using nonlinear least squares fitting.  The polynomial form with relatively few coefficients 
($n = 5$ in this work) is warranted by the approximate linearity of $f(x)$ near $x = 0$.  

\begin{figure}
\includegraphics[width=.4\textwidth]{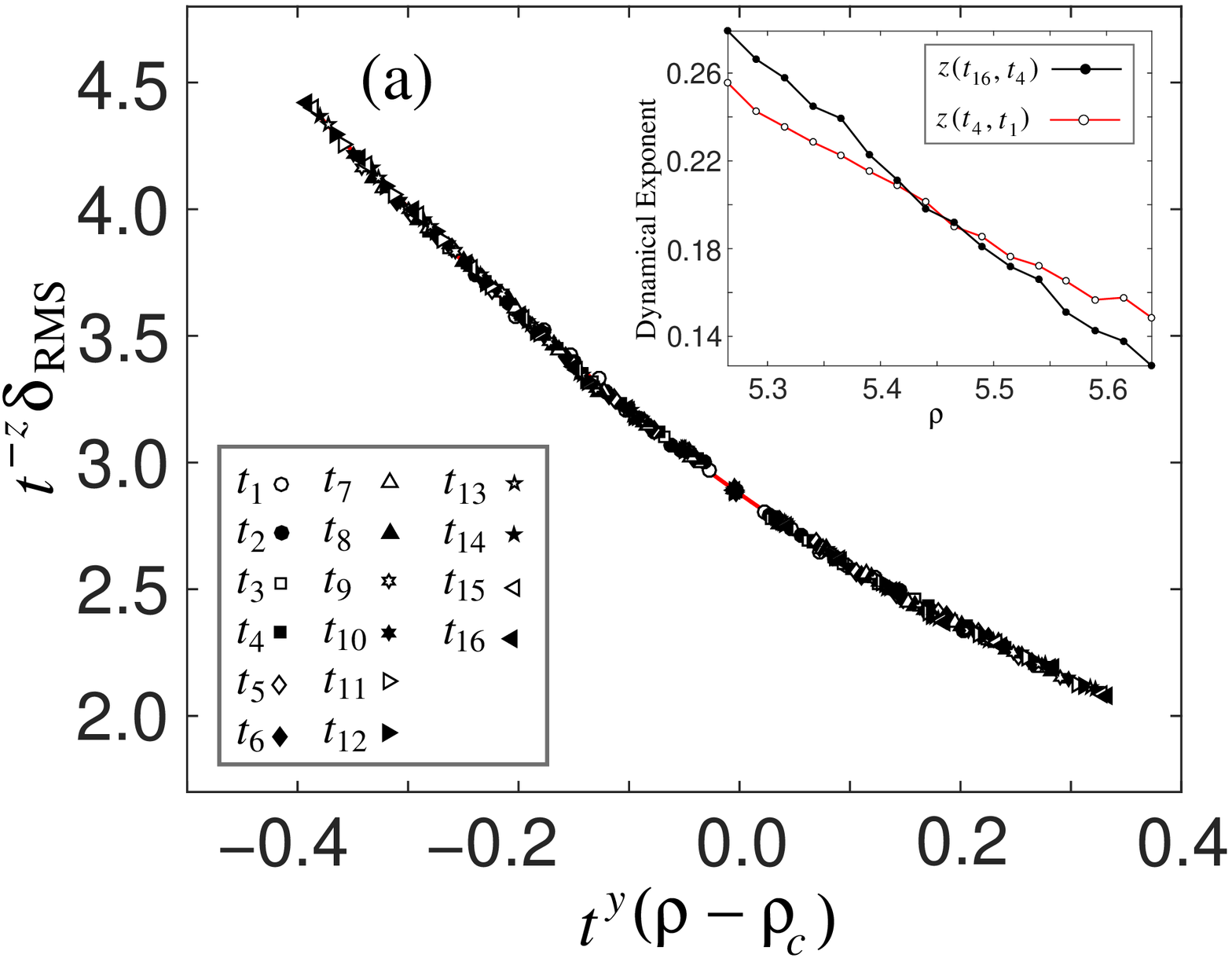}
\includegraphics[width=.4\textwidth]{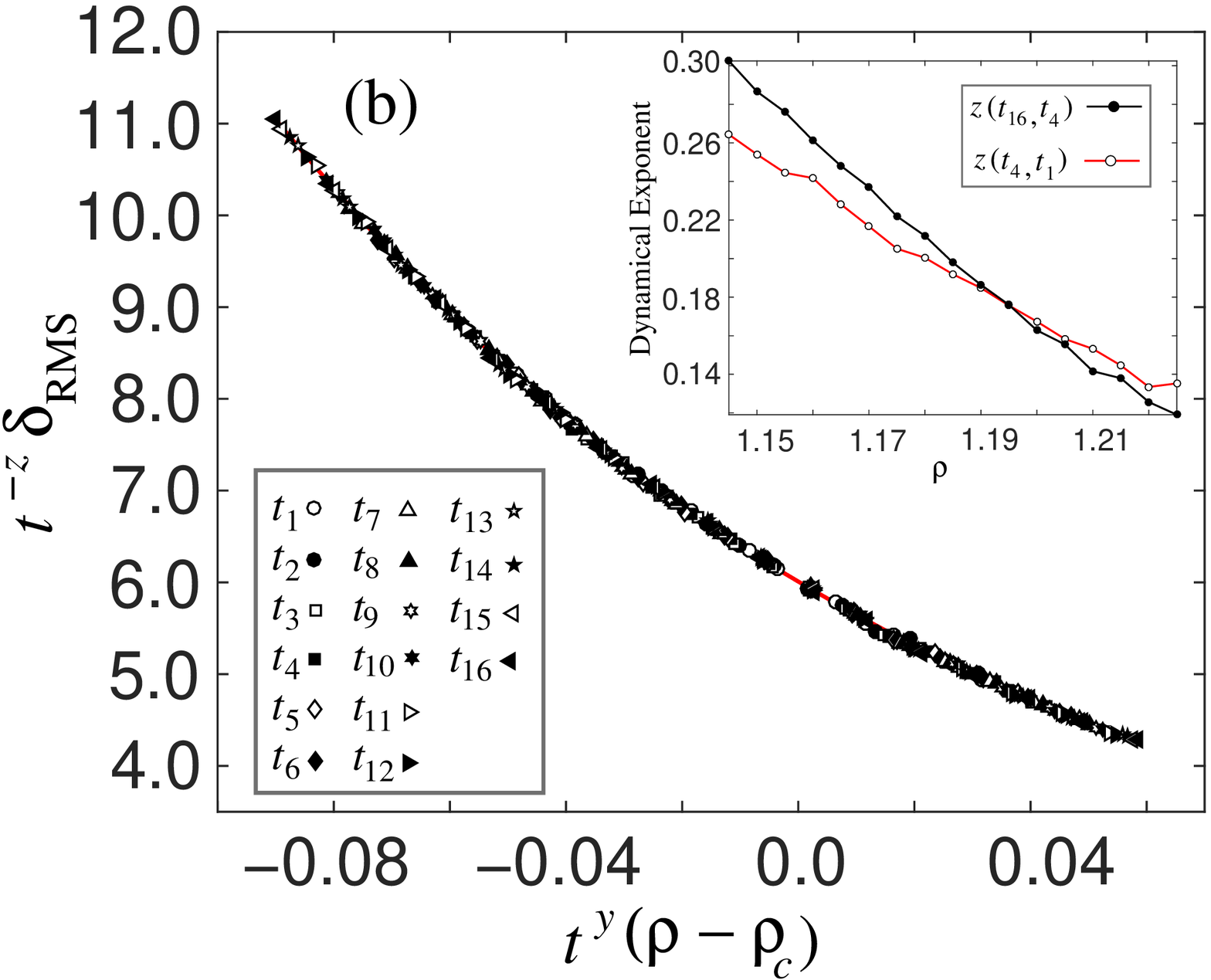}
\caption{\label{fig:Fig4} (Color online) Data Collapse plots are shown for randomly oriented tetrahedra and randomly oriented 
dodecahedra in panel (a) and panel (b), respectively (red lines represent analytical scaling curves). Legends in the plots show symbols corresponding to 
the 16 subintervals of the total dwell time.  The insets in each graph show the intersection of effective exponents $z(t_{16},t_{4})$ and 
$z(t_{4},t_{1})$.}
\end{figure}

Sample data collapses are shown in panel (a) and panel (b) of Fig.~\ref{fig:Fig4} 
for randomly oriented tetrahedra and dodecahedra, respectively.
Data collapses yield the percolation threshold concentrations $\rho_{c}$ to within a few tenths of a percent, as 
well as the dynamical critical exponent $z$ and the scaling exponent $y$.

The graph insets show crossings of effective dynamical exponents $z(t_{16},t_{4})$ and $z(t_{4},t_{1})$ where
$z(t_{2},t_{1}) \equiv \ln[\delta_{\mathrm{RMS}}(t_{2})/\delta_{\mathrm{RMS}}(t_{1})]/\ln(t_{2}/t_{1})$.
Critical densities obtained from the crossings are in agreement with 
data collapse results up to Monte Carlo error.

\begingroup
\squeezetable
\begin{table}[h]
\centering
\begin{tabular}{| c| c | c | c | c | c|}
\hline
Grain Type & $\rho_{c}$  & $v_{\mathrm{B}}$ & $\phi_{c}$  & $z$ & $y$ \\
\hline
\hline
Aligned Tetrahedra & 5.47(2)  & 0.5132 & 0.0605(6) & 0.20(1) & .23(1) \\
\hline
Rotated Tetrahedra & 5.47(2)  & 0.5132 & 0.0605(6) & 0.19(1) & .24(1) \\
\hline
\hline
Aligned Cubes & 2.122(5)  & 1.5396  & 0.0381(3) &  0.20(1) &  .25(1) \\
\hline
Rotated Cubes & 2.012(8) & 1.5396 & 0.0452(6)  & 0.18(1) &  .23(1) \\
\hline
\hline
Aligned Octahedra & 2.402(6)  & 1.3333 & 0.0407(3) &  0.19(1) & .25(1) \\
\hline
Rotated Octahedra & 2.419(9) & 1.3333 & 0.0398(5) & 0.19(1) & .23(1) \\
\hline
\hline
Aligned Dodecahedra & 1.198(3) & 2.7852 & 0.0356(3) &  0.18(1) & .24(1) \\
\hline
Rotated Dodecahedra & 1.194(4) & 2.7852  & 0.0360(3) & 0.18(1) &  .22(1) \\
\hline
\hline
Aligned Icosahedra & 1.326(3) & 2.5362 & 0.0346(3) &  0.19(1) & .24(1) \\
\hline
Rotated Icosahedra &  1.338(8) & 2.5362 & 0.0336(7) & 0.18(1)  & .23(1) \\
\hline
\hline
Spherical Grains & 0.837(1) & 4.1888  & 0.0301(1) & 0.17(1) & .22(1) \\
\hline
\end{tabular}
\caption{\label{tab:Tab1} Table with critical indices for randomly oriented and aligned Platonic solids as well as spheres.}
\end{table}
\endgroup

Results for randomly oriented Platonic solids and the aligned counterparts are summarized in Table~\ref{tab:Tab1}.
In discussing void percolation phenomena, one often specifies the excluded volume $\phi_{c} = e^{-\rho_{c} v_{\mathrm{B}}}$ 
with $v_{\mathrm{B}}$ being the
inclusion volume.  Standard geometric arguments give in terms 
of the radius of the circumscribed sphere $v_{\mathrm{T}} =  \frac{8 \sqrt{3}}{27}$, 
$v_{\mathrm{C}} = \frac{8 \sqrt{3}}{9}$, $v_{\mathrm{O}} = \frac{4}{3}$, $v_{\mathrm{I}} = \frac{2 \sqrt{3}}{9}(5 + \sqrt{5})^{1/2}$, and 
$v_{\mathrm{D}} = \frac{2 \sqrt{2}}{3}(5 + \sqrt{5})$ 
for the tetrahedron, octahedron, cube, icosahedron, and dodecahedron respectively.

Accordingly,  the critical density $\rho_{c}$  (in units of $R^{-3}$)
and the threshold excluded volumes $\phi_{c}$ are specified for aligned and randomly oriented grains.
One might anticipate 
orientational disorder to be disruptive to void networks and thus lower the percolation threshold. 
However, the results are only partly consistent with this intuition.
Introducing random orientations in the case of cube-shaped grains reduces $\rho_{c}$ by 5\% from 2.122(5) to 2.012(8).  
The $\phi_{c} = 0.0381(1)$ result for assemblies of aligned cubes is nearly within error of the value $\phi_{c} = 0.36(1)$ reported previously~\cite{Koza}.
Nevertheless, up to Monte Carlo error, we 
do not find distinct percolation thresholds for aligned and randomly oriented configurations for 
non-cubic grains.  In spite of the robustness of $\rho_{c}$ under these circumstances,
the fact that critical densities even for 
dodecahedra and icosahedra  differ significantly from $\rho_{c}$ for spherical 
grains indicates that all of the polyhedra we consider are distinguished in our results from spherical inclusions.
On the other hand,
the prevalence of adjacent parallel non-penetrating 
faces in the case of aligned cubes, which is not replicated in aligned non-cubic polyhedra, could account for the greater 
likelihood of intact void networks than in randomly oriented configurations; alignment of non-cubic inclusions may be of 
less benefit in mitigating the truncation of connected void spaces.

\begin{figure}
\includegraphics[width=.4\textwidth]{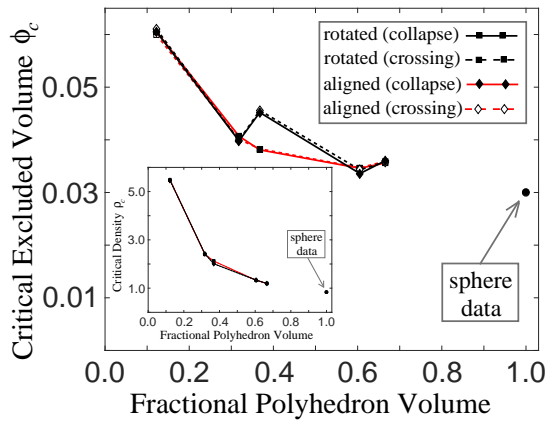}
\caption{\label{fig:Fig5} (Color online) The main graphs shows critical excluded volumes $\phi_{c}$ with respect to the polyhedron volume
fraction relative to that of the circumscribed sphere, and the inset graph displays the threshold density $\rho_{c}$ versus the
polyhedron volume fraction.  The legend corresponds to both the main graph and the inset plot.}
\end{figure}

The critical indices in Table~\ref{tab:Tab1} include dynamical exponents $z$ and scaling exponents $y$;  
up to Monte Carlo error 
the results are compatible with a candidate universal dynamical exponent $z = 0.19(1)$, which is also in
accord with a $z$ obtained in the context of a 3D discrete system~\cite{Stauffer}.

Threshold densities are represented graphically in Fig.~\ref{fig:Fig5} with $\phi_{c}$ and $\rho_{c}$ data in 
the main graph and inset, respectively; diamonds or squared represent results for polyhederal inclusions 
per the main graph legend, while round symbols 
indicate spherical grains.  Apart from cube-shaped grains, $\phi_{c}$ and $\rho_{c}$ data match closely.  The threshold density in 
the inset is monotonic in increasing polyhedral volume fraction, a trend slightly disrupted for the excluded volume, where $\phi_{c}$ 
decreases until rising slightly among icosahedra and dodecahedra.

In conclusion, we have calculated percolation thresholds $\rho_{c}$ for randomly oriented interpenetrating polyhedral 
grains for each of the Platonic solids, the first calculation of critical densities and indices 
characterizing critical behavior for percolation phenomena through voids around randomly oriented polyhedra. The 
diminished $\rho_{c}$ for randomly oriented cubes relative to aligned counterparts is 
not replicated for the other solids considered.  Results for the dynamical critical exponent $z$ are consistent with  
a $z = 0.19(1)$ candidate for a universal value.

\begin{acknowledgments}
We acknowledge helpful discussion with Michael Crescimanno.
Calculations in this work have benefitted from use of the Ohio Supercomputer facility (OSC)~\cite{OSC}.
\end{acknowledgments}


\end{document}